\documentclass[12pt, letter, twocolumn]{IEEEtran}

\usepackage{graphicx}
\usepackage{epsfig}
\usepackage{algorithm}
\usepackage{algorithmic}
\usepackage{ifmtarg}
\usepackage{caption}
\usepackage{footnote}
\usepackage{cite}
\usepackage{amssymb}
\usepackage{amsthm}
\usepackage[fleqn]{amsmath}
\usepackage{amsmath}
\usepackage{epstopdf}
\usepackage{amsfonts}
\usepackage{enumitem}
\usepackage{subfigure}
\usepackage{multirow}
\usepackage{rotating}
\usepackage{hyperref}
\usepackage{color}
\usepackage{comment}

\usepackage[font=normalsize]{caption}
\begin{document}
\title{Facilitating Satellite-Airborne-Terrestrial Integration for Dynamic and Infrastructure-less Networks}
\author{\IEEEauthorblockN{Ahmad Alsharoa, \textit{Senior Member, IEEE}, and Mohamed-Slim Alouini, \textit{Fellow, IEEE}}
\thanks { \vspace{-0.5cm}\hrule
\vspace{0.1cm} \indent Ahmad Alsharoa is with the Electrical and Computer Engineering Department, Missouri University of Science and Technology, Rolla, Missouri 65409, USA, E-mail: aalsharoa@mst.edu.
\newline \indent Mohamed-Slim Alouini is with the Computer, Electrical and Mathematical Sciences and Engineering (CEMSE) Division, King Abdullah University of Science and Technology (KAUST), Thuwal, Makkah Province, Saudi Arabia. E-mails: slim.alouini@kaust.edu.sa.
}\vspace{-.1cm}}

\maketitle
\thispagestyle{empty}
\pagestyle{empty}

\begin{abstract}
\boldmath{
This paper studies the potential improvement in the achievable data rate available to ground users by integrating satellite, airborne, and terrestrial networks. The goal is to establish dynamic wireless services in remote or infrastructure-less areas.
This integration uses high-altitude platforms in the exosphere, stratosphere, and troposphere for better altitude reuse coupled with emerging optical or other high-frequency directional transceivers. Hence they offer a significant increases in the scarce spectrum aggregate efficiency.
However, managing resource allocation with deployment in this integrated system still has some difficulties.
This paper aims to tackle resource management challenges by (i) providing wireless services to ground users in remote areas and connecting them with metropolitan and rural areas,
(ii) employing high-altitude platforms (HAPs) equipped with free-space-optical communication modules for back-hauling backbone.
Finally, we show how our results illustrate the advantages of using the proposed scheme.
}
\end{abstract}


%\section{Introduction}
\section{Introduction and Motivation}\label{Introduction}

Satellite and terrestrial stations are currently the main communication systems that provide wireless services to ground users in remote and metropolitan areas, respectively. While traditional space communications, including satellite stations, are able to deliver broadband services to ground users in remote areas, the spectral efficiency is constrained due to high path-loss attenuation of the communication links between the satellite stations and the ground users.
On the other hand, terrestrial stations cannot support ground users in remote areas due to their limited coverage areas.
Integrating the terrestrial network with satellite stations is one of the proposed solutions to increase the network's coverage and capacity.
Satellite stations can use multiple spot beams associated with multiple protocol label switching and spectrum access control~\cite{sat1}. Also, satellite stations can communicate with users with the help of terrestrial base stations working as relays~\cite{sat1}. In this case, terrestrial base stations are used to amplify the communication links between multiple satellite stations and multiple users~\cite{sat4}.
However, most of traditional satellite communication techniques are constrained due to high path-loss attenuation of the communication links between the satellite stations and the ground users. Also, satellite stations can be located at various orbital heights, thus causing extra delay when providing real-time services to ground users.
Furthermore, satellite stations depend on the terrestrial network to broadcast their signals, which presents another limitation especially in remote areas, and during periods of congestion or network failure.

\noindent\textbf{High-altitude platforms (HAPs) can fill this gap and provide downlink services in remote and congested areas:}
One of the most important elements in the 6G network is that the coverage needs to be large enough to provide acceptable data communications services wherever the users are living, including urban and remote areas. 6G networks are not intended to provide equally good service to all areas but rather maintain some resource balance~\cite{6G}.
With the rapid growth in mobile and wireless devices usage in addition to huge data traffic, traditional terrestrial networks are expected to face difficulties in supporting the demands of users in urban areas. The problem can be exacerbated if failures occur in the ground infrastructures.
On the other hand, it is infeasible to develop terrestrial infrastructure that provides telecommunication services to remote areas.
To mitigate these limitations, integrating the current terrestrial network with higher altitude stations has become a promising technique to provide global connectivity.

To achieve this communications goal, a system of HAPs, such as airships operating in the stratosphere, at altitudes of 17 Km to 20 Km, have been proposed as a wireless solution that can be integrated into existing satellite and terrestrial networks and, hence, improve the overall network throughput. This altitude range is chosen because of its low wind currents and low turbulence, which reduces the energy consumption needed to maintain the position of the HAPs~\cite{Airborne_survey}. The HAPs can act as aerial base stations to improve the communication links between satellite stations and ground users. Hence, HAPs can enhance the overall network throughput and help global connectivity with or without the existence of a terrestrial network~\cite{HAP2}.
This solution can provide immediate wireless connectivity to
(i) ground users in remote areas with infrastructure-less networks,
(ii) on-demand users in congested urban areas experiencing capacity shortage due to peak traffic, e.g., Olympic games, marathons, or base stations failure,
(iii) first responders and victims in an emergency or disaster-recovery situation where the infrastructure network is unavailable or disrupted,
(iv) ground military in a hostile environment,
and (v) border patrol services for patrolling in a difficult terrain.
The main advantages of using HAPs over terrestrial stations can be summarized as follows~\cite{Airborne_survey,HAP2}:
\begin{itemize}
  \item High coverage area: The broadband coverage area of terrestrial stations is usually limited compared to HAPs due to high non-line-of-sight pathloss.
  \item Dynamic and quick deployment: HAPs have the ability to fly to infrastructure-less regions to provide on-demand services.
  \item Low energy consumption: HAPs can be equipped with solar panels that collect energy during the daytime. Thus, with a careful trajectory optimization, HAPs can be self powered~\cite{fb_HAP}.
\end{itemize}
On the other hand, the main advantages of using HAPs over the satellite stations can be summarized as follows~\cite{Airborne_survey}:
\begin{itemize}
  \item Quick and low cost deployment: HAPs can accommodate temporal and traffic demand quickly, where one HAP is sufficient to start providing the service. Also, HAPs can play a significant role in emergency or disaster relief applications by flying to desired areas, in a short and timely manner, in order to restart communications. In addition, the deployment cost of HAPs is much less than satellite deployment costs.
  \item Low propagation delays and strong signals: Due to the high path-loss attenuation between satellite stations and ground users, HAPs can provide services to ground users with less delay.
\end{itemize}

Several papers in the literature have studied the deployment of HAPs~\cite{HAP3,fb_HAP,fb2_HAP}. For instance, the work proposed in~\cite{HAP3} investigates HAPs deployment while taking into consideration the quality-of-service (QoS) to ground users. The authors propose a self-organized game theory model, where the HAPs are modeled as rational and self-organized players with the goal of achieving the optimal configuration of HAPs that maximizes the users' QoS. Furthermore, the work in~\cite{fb_HAP} proposes trajectory optimization techniques for HAPs equipped with solar power panels, employing a greedy heuristic and realtime solution to optimize the HAPs' trajectory by minimizing the consumed energy, which is constrained by the amount of the harvested energy. In~\cite{fb2_HAP}, the authors expanded on their work presented in~\cite{fb_HAP} to include several trajectory optimization methods to maximize the storage energy in HAPs instead of minimizing the consumed energy.
Furthermore, improving the system throughput is considered to be another important key factor in HAPs communications. The authors in~\cite{HAP4} propose a multicast system model that uses orthogonal frequency division multiple access (OFDMA) to find the best transmit powers, time slots, and sub-channels to maximize the total ground users' throughput. In other words, they maximized the number of users that receive the requested multicast streams in the HAP service area in a given OFDMA frame. The improvement achieved in the multiple HAPs' capacity is presented in~\cite{HAP5}, where the authors show that HAPs can offer a spectrally efficiency by exploiting the directionality of the user antenna. In other words, the authors explain how multiple HAPs can share the same frequency band by taking advantage of the directionality of the users' antenna. In~\cite{HAP6}, the authors investigate the multi-user multiple-input multiple-output (MU-MIMO) in HAP communications, where the HAPs are equipped with large-scale antenna arrays. They aimed to formulate and solve a low computational complexity technique that maximizes the signal-to-interference-noise ratio (SINR) and limits the interference between users.

\noindent\textbf{Why equipping HAPs with free-space-optical (FSO) transceivers is a tipping point:}
There is limited research in the literature that has proposed equipping HAPs with FSO transceivers~\cite{FSO10,FSO17}. For instance, in~\cite{FSO10}, the authors provide an overview of HAPs equipped with optical transceivers and show that by using leaser beams, a data rate of several Gbps can be achieved. While in~\cite{FSO17}, a closed form expression for bit-error-rate and average capacity is derived for
multi-hop FSO links in the stratosphere.
However, all these works have not considered managing the resource allocation in satellite-airborne-terrestrial network integration and maintenance of the FSO communication links. In addition, the previous works have not considered issues surrounding access and back-hauling links communications.

\section{HAPs-relay integration with hybrid RF/FSO}

\begin{figure}[t!]
\includegraphics[width=3.5in]{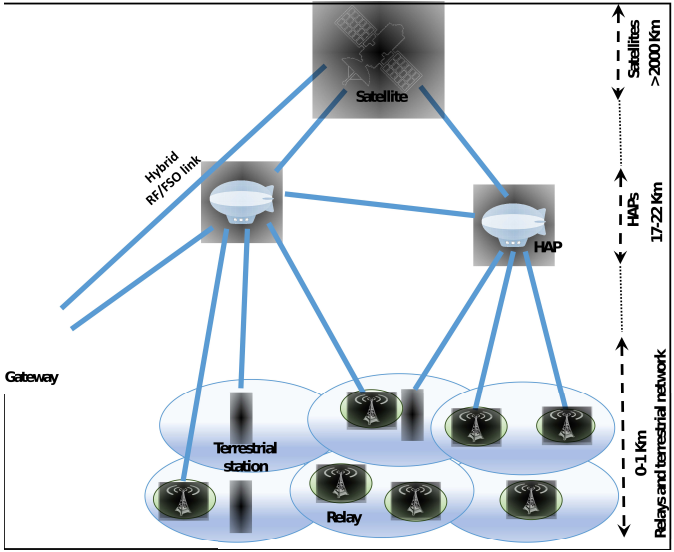}
\caption{System model.}\label{SystemModel_u2}
\end{figure}
%\noindent \textbf{Can HAPs alone do downlink and uplink communications?}
Some large geographical areas may require HAP stations to provide downlink wireless service, but due to power limitation of the ground users, there will be a need for intermediate nodes such as a relay, which broadcast and/or amplify the uplink signals. Relays not only improve the downlink/uplink signal, but also provide a new dimension to the next-generation wireless networking and service provisioning.
For example, in 2017, Hurricane Irma damaged a significant proportions of the wireless cell towers in several parts of Florida. Even after  2 days following the hurricane, Irma’s wrath had caused more than 50\% cell tower failure in some counties in Miami, FL.%~\cite{FCCirma}.
In such circumstances, HAPs and relays are capable of reaching such affected areas thanks to their quick and dynamic deployment.%~\cite{2017-Stumpe-drones}.
%Furthermore, the airline industry embraces LAPs/relays as cost-saver that enable video network feeds resulted in search and rescue of people \cite{2017-Stumpe-drones}.
The success of deploying relays in remote or challenging areas does not depend only on their integration with ground users through the access links, but also on other parameters related to the back-hauling links.%\cite{2018-Rhode-drone}.
Therefore, the integration between HAPs and relays technology with highly efficient placement and resource managements is a great way to solve wireless connectivity in challenging areas. In this study, we propose that the relays are powered by renewable energy (RE) sources to serve users in remote areas. We consider this a valid and plausible proposal since in remote areas the relays will serve fewer users and, hence, will not consume as much energy as when broadcasting the signals.

%\subsection{HAPs-relay integration with hybrid RF/FSO}
Prospective demands of next-generation wireless networks are ambitious and will be required to support data rates 1000 times higher and round-trip
latency 10 times lower than current wireless networks. The radio frequency (RF) spectrum is expected to become more congested for emerging applications in
next-generation wireless communications and eventually insufficient to accommodate the increasing demand of mobile data traffic.
Relying on improved RF access in legacy bands alone is not sufficient; thus, it is critical to embrace a comprehensive solution with \emph{high spectral reuse} by supplementing RF with other emerging wireless technologies in directional high-frequency bands~\cite{2013-Sevincer-LIGHTNETS}.
FSO communications present a promising complementary solution to meet the exploding demand for wireless communications. FSO transceivers are amenable to dense integration, can be modulated at high speeds and provide spatial reuse/security through highly directional beams. RF's unlicensed huge spectrum presents a great opportunity for future spectrum-scarce mobile networks. Its potential integration with solid-state lighting technology presents an attractive commercialization possibility in the long run~\cite{2013-Sevincer-LIGHTNETS}. The authors in~\cite{FSO_5G} proposed a vertical framework consisting of networked HAPs that support back-hauling links and access links of small cell base stations in a multi-tier heterogeneous network. However, that work is limited to supporting small cells and does not consider the integration of all types of terrestrial base stations (e.g., macro cell base station) and satellite stations with HAPs. In addition, that work does not discuss providing global connectivity to remote areas.
%This paper focuses on infrastructure-less operation, where the HAPs and RE relays are equipped with directional FSO transceivers to provide wireless connectivity for the back-hauling links.

\begin{figure}[t!]
\includegraphics[width=3.5in]{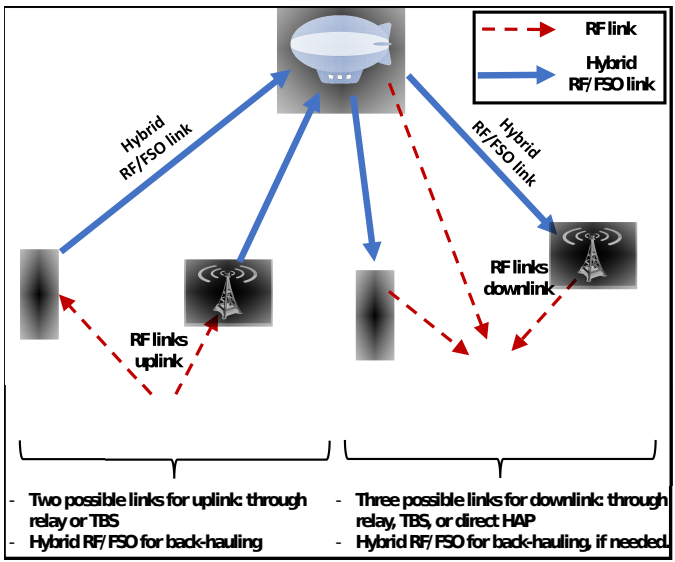}
\caption{All possible uplinks and downlinks scenarios}\label{4scenarios}
\end{figure}

HAP-relay integration offers another dimension to legacy wireless networking by enabling spatial reuse. The relay station can effectively amplify the signals between the HAPs and ground users using an FSO link without causing any major interference to the rest of the ground users. If this spatial technique is not used and relaying is used only on the RF band, then the aggregated throughput will be limited due to (i) the scarcity of the RF band and (ii) the possibility of interference between HAPs, relays, and ground users.
%Therefore, the potential of using spatial reuse is possible only if HAPs, relays, and gateways are equipped with FSO directional antennas and their positioning and precision steering of the antennas are feasible~\cite{2011-singh-interference,2010-sani-directional,2009-ramachandran-r2d2,2014-khan-maintaining}.
%However, the highly directional FSO transceivers require establishment and maintenance of line-of-sight (LoS) between transceivers. Figure~\ref{4scenarios} shows the four possible transmission scenarios of HAPs-relay integration using different access and back-hauling links.
The use of a hybrid RF/FSO link will be for back-hauling links only. The RF and FSO choice will depend on several factors such as environmental/weather conditions and the feasibility of LoS. Meanwhile, the RF band can be used in the access link due to the difficulties in tracking the movement of ground users and maintaining the LoS.

\section{System Model}
We define the following system model as a
\begin{itemize}[leftmargin=*]
\item
Set of stations including a satellite, HAPs, relays, and terrestrial stations,
where each station contains the station ID, 3D location, battery level at any one time (if applicable, e.g., HAPs and relays), and back-hauling rate. All this information is shared with some central units. Since the relays are RE stations, then the consumed power of the relays such as operating and transmission are included.
\item
Set of ground users that contains the user ID, 2D location, and QoS requirement.
\item
Set of fix gateways that contains the gateway ID and 2D location.
\end{itemize}

The major challenge is managing the resources of the ground users and stations in order to maximize the users' data rate utility, taking into consideration the bandwidth and power constraints, association constraints, and placement constraints. Note that the throughput depends on several factors such as the maximum allowable power of the users and HAPs, maximum bandwidth, and HAPs' placement.
A control link between the HAPs and corresponding users is required, i.e., a control link, so that the HAPs can keep track of the users under its coverage area.
The rate utility of the system can be characterized in different utility metrics. The selection of the utility metric can be based on the required fairness level. Some examples are (i) sum rate utility (maximize the sum rate throughput of all users), (ii) minimum rate utility (maximize the minimum user throughput), and (iii) proportionally fair rate utility (maximize the geometric mean of the data rate).

\section{Resource Management}
The key  goal is to attain high throughput of ground users and energy efficiency. Several metrics can be implemented to achieve this goal, such as (1) minimizing the total consumed energy while satisfying a certain user's throughput, or (2) maximizing the energy efficiency utility.
In this context, several high-level research questions need to be addressed:
\begin{itemize}
    \item \textit{Resource Optimization}: How to optimize the transmit power allocation of the users, relays, and  HAPs? In addition, given a certain available bandwidth, how to allocate this bandwidth for the control links (for management) or serving links (i.e., access link and back-hauling link)?
    \item \textit{Associations}: Two associations need to be considered, (a) access link association: the association between users and terrestrial stations, relays, HAPs, (b) back-hauling link association: the association between terrestrial stations and relays with HAPs, and between HAPs and gateways, as shown in Fig.~\ref{4scenarios}.
    \item \textit{HAPs Placement}: How to find the best HAPs' locations taking into consideration the back-hauling link quality between terrestrial stations and relays with HAPs.
    \item \textit{FSO Alignment}: How to optimize the FSO alignment angles between different FSO transceivers.
\end{itemize}

\subsection{Access Link Optimization}
%\noindent \textbf{\underline{Task I.A: Using Static LAPs for Given Back-hauling Constraints.}}
In this work, we propose to use multiple stations (i.e., terrestrial stations, RE relays, and HAPs) to provide wireless connectivity to multiple ground users. The access link can be established either directly from HAPs to user or via terrestrial network or relays, given that HAPs are involved in back-hauling, as shown in Fig.~\ref{4scenarios}.
In this case, the mathematical formulation should include an access link binary variable, to indicate that ground users are associated with certain stations for the transmission.
For simplicity, we assume that each user can be associated with one station at most; on the other hand, each station can be associated with multiple users.
For station peak power and users' peak power, an optimization problem can be formulated that maximizes the user' utility given the following constraints: (1) back-hauling bandwidth and rate, (2) stations and users' peak powers, (3) access link associations, and (4) users' QoS. Therefore, the following parameters can be optimized in order to achieve the best objective function: i) transmit power levels of the users' and stations' transmission power, ii) bandwidth allocation to each user, and iii) access link associations.
Note that the back-hauling rates can play a significant role in determining the associations between stations and user.
For example, a user can be associated with a faraway station if it has a good back-hauling rate rather than being associated with a nearby station with a very low back-hauling rate.
Because the relays are RE battery powered devices, then the amount of stored energy by each relay at the end of a given time slot is considered and additional battery limitation constraints on the relays should be respected to ensure that the consumed energy is less than the stored energy in the previous time slot.

%to $C_l(t)=C_l(t-1)+ E^{ch}_l(t)-E^c_l(t)$. Therefore, the following battery constraints must be respected
%\begin{align}
%&E^c_l(t) \leq C_l(t-1), \quad C_l(t-1)+ E^{ch}_l(t) \leq \bar{S}, \quad \forall l, \forall t, \label{power_rbattery}
%\end{align}
%where $\bar{S}$ is the maximum energy that can be stored in the relay. $E^c_l(t)$ and $E^{ch}_l(t)$ are the consumed and charging energies of relay $l$ at the end of time slot $t$.
%The first part of constraint~\eqref{power_rbattery} to ensure that the consumed energy is less than the stored energy in the previous time slot, while the second part of constraint~\eqref{power_rbattery} is to ensure that the charging energy added to previous stored energy should not exceed the relay battery capacity.

\subsection{Back-hauling Optimization}
In this section, we propose the integration of terrestrial stations and relays with HAPs using hybrid RF/FSO links. A key challenge for networking under partial or no infrastructure is to establish a reliable back-hauling links to the relays or gateways that are involved in the provisioning of connectivity services of ground users.
The back-hauling optimization problem is proposed to optimally find back-hauling associations, HAPs' locations, transmit powers, and FSO alignment between transceivers in order to maximize the users' back-hauling throughput while respecting the resource limitations.
This involves utilizing high-frequency directional bands such as an optical band as much as possible to minimize the interference between transceivers.
Therefore, a binary decision variables needs to be introduced for the back-hauling association with terrestrial stations, relays, or HAPs.
Without loss of generality, we assume that each HAP should be strictly associated with one station for back-hauling (either a gateway or a satellite station). Also, the maximum number of HAPs that can be associated with the same station is limited.
For the FSO link between different transceivers and gateways, we assume that the alignment angle is optimized in order to achieve LoS alignment. In this case, we propose the FSO link discovery and establishment. One way is to explore the out-of-band techniques where support from an RF link is available to exchange the angle and direction. %Furthermore, we propose that the FSO transceivers are equipped with multiple FSO transmitters and have the option of electronically steering the link to another transceiver.

A key problem to solve is how the terrestrial stations, relays, and HAPs will discover each other given an external location server/support like GPS. If no location information is given but RF communication capability is available, then the transceivers can use the RF link to exchange information about how they are oriented.
%One potential solution is to start with a large divergence angle and quickly identify the region the other node resides. Then, iteratively use a transceiver with smaller divergence angle to narrow the region where the other is residing. The downside of this approach is that it requires either a transceiver with an adaptive divergence angle or multiple transceivers with different divergence angles.

\section{Results and Discussion}
We provide the numerical results to outline the benefits of using satellite, airborne, terrestrial networks to improve global connectivity.
The simulation results are set within an area of  180 Km $\times$ 180 Km. Within this area, $U$ users are distributed in three different subareas (i.e., subarea 1: 30 Km$^2$, subarea 2: 30 Km$^2$, and subarea 3: the remaining Km$^2$) with different density distributions. Subarea 1 contains nine terrestrial stations with coverage x: (75 Km to 105 Km), and y: (0 Km to 30 Km) and contains 40\% of the total number of users. Subarea 2, has no terrestrial stations with coverage x: (75 Km to 105 Km), and y: (150 Km to 180 Km) and contains 30\% of the total number of users.

\begin{figure}[h!]
\includegraphics[width=3.5in]{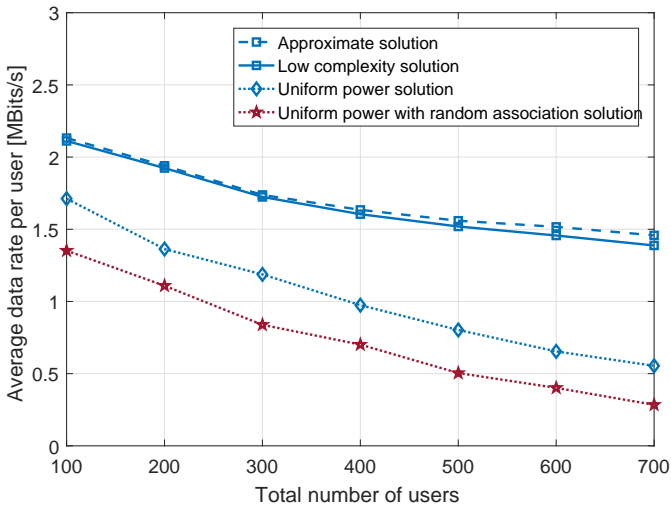}
\caption{Average data rate per user as a function of number of users.}\label{fig1}
\end{figure}

We study the enhancement of the achievable users’ throughput when assisted by terrestrial stations and HAPs. We consider different back-hauling bandwidth cases to represent both RF only and hybrid RF/FSO scenarios in order to investigate the limitation of the RF only scenario.
Fig.~\ref{fig1} plots the average data rate per user (i.e., total sum rate over the number of users) versus total number of users.
We compare our proposed solutions (i.e., approximated and low complexity solutions) with two benchmark solutions:
1- optimizing only the access and back-hauling associations and the HAPs' locations with uniform power distribution (i.e., no power optimization),
and 2- optimizing only the placement of the HAPs using random access and back-hauling links associations with uniform power distribution.
The figure shows that for fixed resources, as the number of users increases, the average data rate decreases due to the limitation of the available resources.
Furthermore, the figure shows that the approximate solution and low complexity solution achieve almost the same performance for a low number of users. However, there is a gap between the two proposed solutions when the number of users is relatively large.

We can explain this by the fact that the low complexity solution forces some HAPs to cover the terrestrial coverage areas when needed. This does not affect the performance when number of users is relatively low, but when the number of users is large, the performance will be affected due to the limited number of HAPs.
Because HAPs have large coverage areas, this gap for a large number of users is still acceptable.
\begin{figure}[h!]
\includegraphics[width=3.5in]{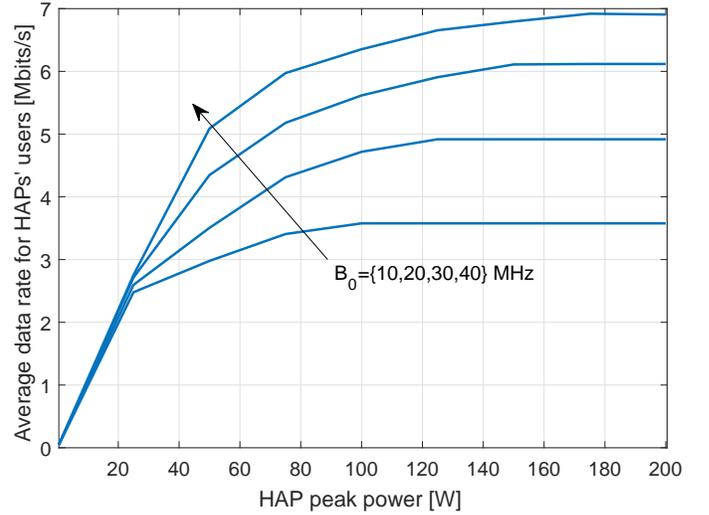}
\caption{Average data rate of HAPs' users versus HAPs' peak power.}\label{fig3P}
\end{figure}

Fig.~\ref{fig3P} shows that our proposed solutions outperform the other two benchmarks solutions. For instance, using number of users$=400$, our proposed solution can enhance the average rate throughput by at least 39\% and 88\% compared to optimizing associations with uniform power, and to random associations with uniform power, respectively.
Furthermore, note that the gap between our proposed solutions and the benchmark solutions increases as $U$ increases. This occurs because, as the number of users increases, the need to manage and optimize the power becomes more necessary.
On the other hand, to illustrate the back-hauling bottleneck, Fig.~\ref{fig3P} plots the average data rate of HAPs' users versus HAPs' peak power.
This figure shows that as the HAPs' peak power increases, the average data rate increases up to a certain value. This can be explained by starting from this value of HAPs' peak power, the average data rate can not be improved because it depends also on
the back-hauling data rate constraint. Furthermore, note that the average data rate increases as back-hauling bandwidth $B_0$ increases,  this occurs because increasing the back-hauling bandwidth also increases the value of the back-hauling data rate, thus increasing the back-hauling bottleneck.
Therefore, hybrid RF/FSO  communication links can be used to mitigate the back-hauling bottleneck limitation, and thus enhance the performance. However, it can add more complexity to the problem by optimizing extra parameters such as the LoS alignment angles.

%%%%%%%%%%%%%%%%%%%%%%%

\section{Conclusion}
This paper proposes an efficient scheme that integrates terrestrial stations, RE relay, HAPs and satellite stations to provide global connectivity.
Our objective is to improve the end-to-end users' throughput by optimizing back-hauling and access links.
In addition, we proposed to equip terrestrial stations, relays, and HAPs with FSO transceivers to mitigate the back-hauling bottleneck limitation, and thus enhance the data rate.
However, this will add more complexity to the problem by optimizing extra parameters such as the LoS angles.

\bibliographystyle{IEEEtran}
%\bibliography{J_2019HAP}
\bibliography{CM_2019HAP_15ref}

\end{document}